\documentstyle[aps,twocolumn,floats]{revtex}
\input psfig.sty
\begin{document}
\title{Crystal Field, Magnetic Anisotropy and Excitations in Rare-Earth
Hexaborides}
\author{Gennadi Uimin}
\address{
Landau Institute for Theoretical Physics, Chernogolovka, 142432 Moscow
Region, Russia}
\author{Wolfram Brenig}
\address{
Institut f\"ur Theoretische Physik, Technische Universit\"at Braunschweig,
38106 Braunschweig, Germany
}
\date{\today}
\maketitle
\begin{abstract}
We clarify the role of crystalline electric field (CEF) induced magnetic
anisotropy in the ground state and spin-wave spectrum of cubic rare-earth
materials with dominating isotropic magnetic exchange interactions.
In particular we study the hexaboride NdB$_6$ which is shown to exhibit
strong spin-quadrupolar coupling. The CEF scheme is analyzed and a
non-collinear magnetization response is found. The spin orientation
in the antiferromagnetically ordered ground-state is identified. Moreover,
the spin excitations are evaluated and in agreement with inelastic
neutron scattering a suppression of one of the two magnetic modes
in the strong-coupling regime is predicted.

\end{abstract}

\subsection*{Introduction}
Over the last two decades cubic rare-earth hexaborides, REB$_6$  (RE,
rare-earth element), with CaB$_6$-type crystal structure have been at
the center of numerous studies of materials with crystalline-electric-field
(CEF) driven, non-trivial ordering phenomena. Among these compounds,
CeB$_6$ (e.g., \cite{effantin}) serves as a prototypical system which
exhibits an impressively complex phase diagram. In this material the CEF
of cubic symmetry selects the $\Gamma_8$ quartet to be the ground state
of the Ce$^{3+}$ ions (${\cal J}=5/2$). The latter quartet is well separated
from the next-highest $\Gamma_7$ doublet by an energy gap of order 540K
\cite{zirngiebl}. Thus, on a low-energy scale, the physics of CeB$_6$
is reasonably well described by projecting onto the $\Gamma_8$ subspace.
Similar systems with $\Gamma_8$ ground states can be realized starting
from the right side of the rare-earth series, i.e., invoking compounds of
cubic symmetry with Yb$^{3+}$ or Tm$^{2^+}$ ions, whose incomplete $f$-shell
contains 13 electron, or one $f$-hole. In accordance with Hund's
rule and contrast to the Ce-case however, the $\Gamma_8$ basis has to
be constructed from a ${\cal J}=7/2$ multiplet, breaking direct
electron-hole symmetry thereby.

In this brief note we will focus on the hexaboride NdB$_6$. Although
investigated in detail experimentally by inelastic neutron scattering (INS)
\cite{loew,erkelens} the anisotropy of the magnetically ordered state below
the temperature $T_C$ of order $T_C\approx 8.6$K \cite{Burlet88,erkelens}
remains unclear as well as the existence
of only a single magnetic mode as observed by INS. The aim of our work is
to consider these open issues.

\subsection*{Crystalline electric field}
The CEF level scheme of the Nd$^{3+}$ multiplet (three $f$-electrons,
${\cal J}=9/2$, $S=3/2$, $L=6$) is consistent with the sequence
$\Gamma_8^{(2)}$(0 K) - $\Gamma_8^{(1)}$(135 K) -
$\Gamma_6$(278 K) \cite{loew}. Similar to CeB$_6$ the energy gap
separating the lowest quartet is large enough to restrict the
Hilbert space to $\Gamma_8^{(2)}$ only. The basis states of the latter
sub-manifold can be represented by the linear combinations \cite{longvers}:
\begin{eqnarray}
\begin{array}{l}
\psi_{+\uparrow}=v_1\left|+9/2\right\rangle +
v_2\left|+1/2\right\rangle + v_3\left|-7/2\right\rangle,\\
\psi_{-\uparrow}=w_1\left|+5/2\right\rangle +
w_2\left|-3/2\right\rangle, \label{spin+_92}
\end{array}
\end{eqnarray}
\vspace{-5mm}
\begin{eqnarray}
\begin{array}{l}
\psi_{+\downarrow}=v_1\left|-9/2\right\rangle +
v_2\left|-1/2\right\rangle + v_3\left|+7/2\right\rangle, \\
\psi_{-\downarrow}=w_1\left|-5/2\right\rangle +
w_2\left|+3/2\right\rangle. \label{spin-_92}
\end{array}
\end{eqnarray}
The coefficients $v_i$ and $w_i$ depend on the respective ratios of the CEF
splittings. For NdB$_6$ one finds \cite{longvers}:
\begin{eqnarray}
\begin{array}{l}
v_1=0.1437,\quad v_2=-0.3615,\quad v_3=0.9212;\\
w_1=-0.9223,\quad w_2=0.3865.
\end{array}
\end{eqnarray}
The states in (\ref{spin+_92}) and (\ref{spin-_92}) have been labeled such
that the 2$^{\rm nd}$ index denotes a "spin"-like
projections, whereas the 1$^{\rm st}$ index stands for two "orbital"-like
components which reflect the different shapes of the electron wave-functions.
This leads to a description of the quartet in terms of two Pauli matrices,
$\bbox{\sigma}$ and $\bbox{\tau}$ \cite{ohkawa,ukf,sst}
\begin{eqnarray}\label{g4}
\sigma^z\psi_{\tau\,\pm}=\pm 1/2\,\psi_{\tau\,\pm},\quad
\sigma^{\pm}\psi_{\tau\,\mp}=\psi_{\tau\,\pm},\nonumber\\
\tau^z\psi_{\pm\,\sigma}=\pm 1/2\,\psi_{\pm\,\sigma},\quad
\tau^{\pm}\psi_{\mp\,\sigma}=\psi_{\pm\,\sigma}.\nonumber
\end{eqnarray}
where the magnetic-moment operator can be
represented in terms of $\bbox{\sigma}$ and $\bbox{\tau}$ by:
\begin{equation}\label{GsM}
M_\alpha=\mu_B\,(\xi+2\eta T_\alpha)\,\sigma_\alpha,\quad (\alpha=x,y,z).
\label{m4}
\end{equation}
Here ${\bf T}$ is a vector with components
\begin{eqnarray}
T_x=-\frac 12\tau_z+\frac{\sqrt{3}}2\tau_x,
\; T_y=-\frac 12\tau_z-\frac{\sqrt{3}}2\tau_x,\;
T_z=\tau_z,
\label{T}
\end{eqnarray}
which transforms according to the $\Gamma_3$ representation.

Note, that for NdB$_6$, $\xi$ and $\eta$ {\em depend} on the parameters of CEF
splitting \cite{longvers} with:
\begin{equation}
\xi = - 0.661 \makebox[5mm][c]{,} \eta = - 6.857
\;\;.
\end{equation}
This identifies NdB$_6$ to be a system with strong coupling of the
magnetic and quadrupolar degrees of freedom.  In the
1$f$-electron(hole) cases $\xi$ and $\eta$ are universal and do {\em
not} depend on the CEF splitting parameters.  For Ce$^{3+}$ $\xi =2$
and $\eta =8/7$, for Yb$^{3+}$ $\xi = -8/3$ and $\eta = -
32/21$. Therefore CeB$_6$ and YbB$_6$ exhibit rather weak
spin-quadrupolar coupling with a characteristic parameter
$\eta/(2\xi)=2/7$.

\subsection*{Exchange anisotropy}
In this section we clarify the spin orientation in the magnetically ordered
ground state. Most likely, the dominant interaction in NdB$_6$ is of
isotropic magnetic exchange type \cite{erkelens}. However, due to the
$\Gamma_8$ ground state a CEF induced magnetic anisotropy exists which {\em
depends on the ratio} $\xi/\eta$. This can be understood easily by
considering the single-ion Zeeman interaction, i.e., $-{\bf H}\cdot{\bf M}$
in an external magnetic field ${\bf H}$. The eigenvalues $\lambda$ in the
$\Gamma_8$-space are
\begin{equation}\label{g8}
\lambda=\begin{array}{c}
\pm\sqrt{\xi^2\!+\!\eta^2\!\pm\!|\eta|\!
\sqrt{(\frac {3\eta^2}2\! -\!2 \xi^2)\! -\!
3 F({\bf n}) (\frac {\eta^2}2\! - \!2 \xi^2)}}
\end{array}
,
\end{equation}
measured in units of $g\mu_B H/2$. This clearly manifests a cubic anisotropy
through the function $F({\bf n})$:
\begin{equation}\label{g9}
F({\bf n})=n_x^4 + n_y^4 + n_z^4, \quad {\bf n}={\bf H}/H
\;\;.
\end{equation}
The anisotropy results in a non-collinearity of the magnetic field
and the magnetization for any general orientation of ${\bf H}$.
Exceptions are the directions $[1 1 1]$, $[1 1 0]$, $[0 0 1]$, and their
crystallographic equivalents. Energetically favorable states are related
either to the cubic axes ($[0 0 1]$-type), if $|\eta|<2|\xi|$, or the cubic
diagonals ($[1 1 1]$-type), if $|\eta|>2|\xi|$. The anisotropy caused by
the CEF disappears, if $|\eta|=2|\xi|$. Therefore, we may conclude
that Ce$^{3+}$, Yb$^{2+}$ and Tm$^{3+}$ $\Gamma_8$ compounds tend to exhibit
"easy axis" anisotropy ($\eta/(2\xi)=2/7$), whereas for Nd$^{3+}$ in NdB$_6$
we have $\eta/(2\xi)\approx 5.19$ which results in "easy diagonal" anisotropy.

Within a mean-field treatment of the exchange interaction
\begin{equation}
-\sum_{{\bf R},{\bf R}'}J_{{\bf R R}'}{\bf S}_{\bf R}\cdot
{\bf S}_{{\bf R}'}
\end{equation}
where $J_{{\bf R R}'}$ is the exchange integral and ${\bf S}_{\bf R}$ the
spin at site ${\bf R}$ the magnetic field in (\ref{g8}) and (\ref{g9}) has
to be replaced by the Weiss field $J_0\langle {\bf S}\rangle/(g\mu_B)$ with
$J_0=\sum_{{\bf R}'}J_{{\bf R R}'}$ if ferromagnetic exchange is dominant.
The Land\'e factor in NdB$_6$ is $g=8/11$. For
bipartite antiferromagnetism (AFM), the Weiss field on sublattice A is
proportional to $-J_0\langle {\bf S}_A\rangle + J_1\langle {\bf S}_B\rangle$
with $J_{0(1)}=(-)\sum_{{\bf R}'} J_{{\bf R R}'}$ for ${\bf R}$ and ${\bf
R}'$ on equal(opposite) sublattices. On sublattice B, one should replace
$A\leftrightarrow B$.

Therefore, in conclusion, we expect $[111]$ orientational ordering in
the ground state of NdB$_6$ if isotropic exchange interactions are
dominant \cite{refnote}.

\subsection*{Magnetic excitations}
In this section we focus on the spin dynamics by considering the
time-dependent magnetic susceptibility
\begin{equation}\label{w1}
\chi^S_{\alpha\beta}({\bf k},t)=i\Theta(t)
\langle [ S_{\alpha{\bf k}}(t), S_{\beta{\bf -k}} ] \rangle
\;\;.
\end{equation}
Lower Greek indices of $\chi$ and the spin operator refer to $x,y,z$ and
boldfaced vectors ${\bf k}$ denote the momentum. We use a spin
operator rescaled by $\eta^{-1}$, i.e., $S_{\alpha{\bf k}}=M_{\alpha{\bf
k}}/(g\mu_B\eta)$. Therefore the dependence
of the magnetic spectrum on the CEF can be expressed solely in terms of the
ratio $\xi/\eta$. To evaluate (\ref{w1}) we proceed via a mean field
analysis consistent with AFM ordering \cite{Burlet88} on a bipartite
lattice.  Rather than employing the Pauli-matrix representation
\cite{ohkawa,ukf,sst} of (\ref{w1}) we perform this analysis using a dyadic
basis to express the spin operator within the $\Gamma_8$ manifold
\cite{MayBeFuldesCEFReview}:
\begin{eqnarray}\label{w2}
&&S_{\alpha{\bf k}}=\frac{1}{\sqrt{2}}S_{\alpha}^{\mu\nu}
(a^{\mu\nu}_{\bf k}+b^{\mu\nu}_{\bf k})
\nonumber \\
&&a^{\mu\nu}_{\bf k}=\sqrt{\frac{2}{N}}\sum_{\bf R}e^{-i\bf kR}
a^{\mu\nu}_{\bf R}
\nonumber \\
&&b^{\mu\nu}_{\bf k}=\sqrt{\frac{2}{N}}\sum_{\bf R'}e^{-i\bf kR'}
b^{\mu\nu}_{\bf R'}
\end{eqnarray}
where a summation over repeated indices is implied for the remainder of this
paper and
\begin{equation}\label{w2a}
a^{\mu\nu}_{\bf R}=|\mu{\bf R}\rangle\langle\nu{\bf R}|
\makebox[.5cm][c]{,}
b^{\mu\nu}_{{\bf R}'}=|\mu{{\bf R}'}\rangle\langle\nu{{\bf R}'}|
\end{equation}
are the dyades on sites ${\bf R}({{\bf R}'})$ of the magnetic
A(B)-sublattice. $|\mu\rangle$ are the eigenstates of the z-component of the
spin in the $\Gamma_8$-manifold
$S_{\alpha=z}|\mu\rangle=s_{\mu}|\mu\rangle$. The spin should be {\em
quantized along(against)} the $[111]$ direction of the Weiss-field on the
A(B)-sublattice sites. $S_{\alpha}^{\mu\nu}$ are the matrix elements of the
spin corresponding to the latter quantization direction. The dyadic
transition operators $a^{\mu\nu}_{\bf k}$ and $b^{\mu\nu}_{\bf k}$ with
$\mu,\nu=1\ldots 4$ can be recast into a 32-component operator
$A^{\gamma=1\ldots 32}_{\bf k}=\{ a^{(1,1),\ldots (4,4)}_{\bf
k},b^{(1,1),\ldots (4,4)}_{\bf k} \}$ with a corresponding 32$\times$32
matrix-susceptibility of the $A^{\gamma}_{\bf k}$ operators
\begin{equation}\label{w5}
\chi^{\mu\nu}({\bf k},t)
= i\Theta(t)
\langle [ A^{\mu}_{\bf k}(t), A^{\nu\;\dagger}_{\bf k} ] \rangle
\;\;,
\end{equation}
The original magnetic susceptibility (\ref{w1}) can be obtained from
this by projecting the dyades onto the magnetic moment
\begin{equation}\label{w15}
\chi_{\alpha\beta}({\bf k},t) =
\frac{1}{2} \chi^{\mu\nu}({\bf k},t) C^{\nu\mu}_{\beta\alpha}
\;\;,
\end{equation}
where $C^{\nu\mu}_{\beta\alpha}= v^{\nu\;\star}_{\beta}
v^{\mu}_{\alpha}$ with
$v^{\mu=1\ldots 32}_{\alpha=x,y,z} =$
$\{S_{\alpha}^{(1,1),\ldots (4,4)},$
$S_{\alpha}^{(1,1),\ldots (4,4)}\}$ is 32-component vector
for each spin component $\alpha$.

To proceed we evaluate the equation of motion (EQM) of the dyadic
susceptibility
\begin{eqnarray}\label{w7}
i\partial_t\chi^{\mu\nu}({\bf k},t)=&&-\delta(t)
\langle [ A^{\mu}_{\bf k}, A^{\nu\;\dagger}_{\bf k} ] \rangle
\nonumber \\
&&+i \Theta(t)
\langle [ [ A^{\mu}_{\bf k}(t), H], A^{\nu\;\dagger}_{\bf k} ] \rangle
\;\;.
\end{eqnarray}
In this brief report we concentrate on the spin dynamics for next-neighbor
(NN) AFM exchange-couplings $J$ only. The effects of longer-ranged couplings
will be discussed elsewhere \cite{longvers}.  Therefore, setting $J\eta^2/g^2$
to unity the Hamiltonian in terms of the dyades reads
\begin{eqnarray}\label{w6}
H = \sum_{{\bf R}, {\bf l}}
S_{\alpha}^{\mu\nu}S_{\alpha}^{\lambda\sigma}
a^{\mu\nu}_{\bf R} b^{\lambda\sigma}_{{\bf R}+{\bf l}}
\;\;,
\end{eqnarray}
where ${\bf l}$ runs over the NN sites of ${\bf R}$. The real-space
representation of the commutator on the r.h.s of the EQM is evaluated using
the algebra of the dyades, yielding
\begin{equation}\label{w8}
[ a^{\mu\nu}_{\bf R},H ] = \sum_{\bf l}
(S_{\alpha}^{\nu\omega}
a^{\mu\omega}_{\bf R}  -
S_{\alpha}^{\omega\mu}
a^{\omega\nu}_{\bf R})
S_{\alpha}^{\lambda\sigma}
b^{\lambda\sigma}_{{\bf R}+{\bf l}}
\;\;.
\end{equation}
An analogous expression results on the B sub-lattice. On the mean-field
level the EQMs are closed by factorizing all quadratic terms in (\ref{w8})
according to the scheme
$a^{\mu\nu}_{\bf R}
b^{\lambda\sigma}_{{\bf R}'} = \langle a^{\mu\nu}_{\bf R}\rangle
b^{\lambda\sigma}_{{\bf R}'} + a^{\mu\nu}_{\bf R} \langle
b^{\lambda\sigma}_{{\bf R}'}\rangle$.
Moreover, 'up'('down') $[111]$-polarization on the A(B) sub-lattice is
enforced by setting
\begin{equation}\label{w9}
\langle a^{\mu\nu}_{\bf R}\rangle = \delta^{\mu1}\delta^{\nu1}
\makebox[.5cm][c]{,}
\langle b^{\mu\nu}_{{\bf R}'}\rangle = \delta^{\mu4}\delta^{\nu4}
\;\;,
\end{equation}
In momentum space the linearization results in
\begin{eqnarray}\label{w9a}
[a^{\mu\nu}_{\bf k},H]&=&z
 S_{\alpha}^{44}
(S_{\alpha}^{\nu\sigma}\delta^{\mu\lambda}-
 S_{\alpha}^{\lambda\mu}\delta^{\nu\sigma})a^{\lambda\sigma}_{\bf k}
\nonumber \\
&&+z \gamma_{\bf k}
(\delta^{1\mu}S_{\alpha}^{\nu1}-
 S_{\alpha}^{1\mu}\delta^{\nu1})
 S_{\alpha}^{\lambda\sigma}b^{\lambda\sigma}_{\bf k}
\nonumber \\
& = & z (L^{\mu\nu\lambda\sigma}_{{\bf k}11} a^{\lambda\sigma}_{\bf k}
+ L^{\mu\nu\lambda\sigma}_{{\bf k}12} b^{\lambda\sigma}_{\bf k} )
\end{eqnarray}
where $z$ is the coordination number and $z \gamma_{\bf k}=\sum_{\bf
l}e^{i{\bf k}\cdot{\bf l}}$. A similar equation arises for $[b^{\mu\nu}_{\bf
k},H]$ introducing two additional 16$\times$16 matrices
$L^{\mu\nu\lambda\sigma}_{{\bf k}22}$ and $L^{\mu\nu\lambda\sigma}_{{\bf
k}21}$. Switching to frequency space the EQMs can be solved as
\begin{equation}\label{w10}
\chi^S_{\alpha\beta}({\bf k},\omega)
= - (\omega\,{\bf 1}-z{\bf L}_{\bf k})^{-1} \mbox{\boldmath{$\chi$}}_0
{\bf C}_{\beta\alpha}
\;\;,
\end{equation}
where boldfaced symbols refer to matrix notation in a 32$\times$32 space.
${\bf L}_{\bf k}$ is set by $L^{\mu\nu\lambda\sigma}_{{\bf k},ij}$ with
$i,j=1,2$ labeling four 16$\times$16 sub-blocks. Similarly
$\mbox{\boldmath{$\chi$}}_0$ consists of four sub-blocks
$\chi^{\mu\nu\lambda\sigma}_{0,ij}$ with
$\chi^{\mu\nu\lambda\sigma}_{0,i\neq j}=0$ and
$\chi^{\mu\nu\lambda\sigma}_{0,11(22)}=
\delta^{\nu\sigma}\delta^{\mu1(4)}\delta^{\lambda1(4)}
-\delta^{\lambda\mu}\delta^{\sigma1(4)}\delta^{\nu1(4)}
$.

\begin{figure}[t]
\vskip -.5cm 
\centerline{\psfig{file=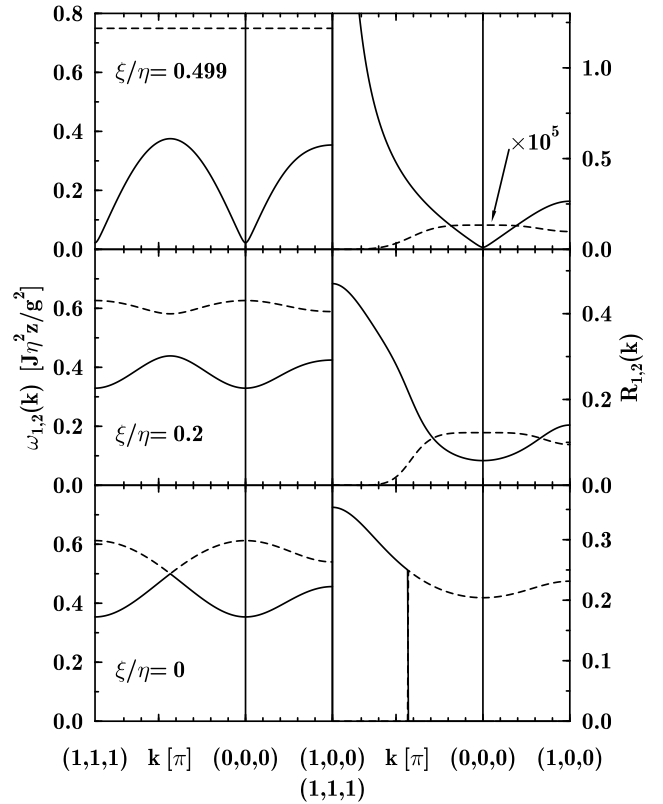,width=9.5cm}}
\vskip -1.7cm 
\caption[l]{Dispersion and weight of spin excitations.}
\label{diszfacfig}
\end{figure}

Eqn. (\ref{w10}) allows for substantial simplifications. First,
all diagonal dyades, i.e. $a(b)^{\mu\mu}_{\bf k}$, commute with $H$.
Second, the linearized form of (\ref{w8}) for the non diagonal dyades,
i.e. for $a(b)^{\mu\nu}_{\bf k}$ with $\mu\neq\nu$, is diagonal with
respect to $\mu\nu$ {\em and} remains local for nearly all pairs
$\mu\nu$. This follows from the identity \cite{longvers}
\begin{equation}\label{w11}
S_{\alpha}^{11(44)} S_{\alpha}^{\mu\nu} =0
\;\;.
\end{equation}
The only set of dyades which couple dispersively via the EQMs is
\begin{equation}\label{w12}
B^{\gamma=1\ldots 4}_{\bf k}=\{
a^{(1,2)}_{\bf k},a^{(3,1)}_{\bf k},b^{(3,4)}_{\bf k},b^{(4,2)}_{\bf k} \}
\;\;,
\end{equation}
and the corresponding hermitian conjugate set $B_{\bf k}^{\gamma=1\ldots 4\;
\dagger}$.
From the preceding discussion it is conceivable that the complete
spin dynamics can be expressed in terms of the physically {\em relevant
dyades} $B^{\gamma=1\ldots 4\;(\dagger)}_{\bf k}$ only. In fact, after some
elementary rearrangements of the matrix-EQM (\ref{w10}), the longitudinal
spin susceptibility, which, due to cubic symmetry, is identical to the
three-trace $\chi^S_{\alpha\alpha}({\bf k},\omega)$ simplifies to
\begin{equation}\label{w13}
\chi^S_{\alpha\alpha}({\bf k},\omega) = -Tr [D^{-1} N]
\end{equation}
where the dynamical matrix $D$ and the static
sus\-cep\-ti\-bi\-li\-ty-ma\-trix $N$ are identical to
$((\omega{\scriptstyle/}z){\bf 1}-{\bf L}_{\bf k})$ and
$\mbox{\boldmath{$\chi$}}_0{\bf C}_{\alpha\alpha}{\scriptstyle/}z$ {\em
restricted} to within the 4 dimensional subspace spanned by eqn.
(\ref{w12}). The complex conjugate dyades $B_{\bf k}^{\mu\;\dagger}$
introduce an overall prefactor of 2 only.  After some algebra we find that
$D$ and $N$ are determined by five parameters $a,b,c,d$, and $e$ through
\begin{eqnarray}\label{w14}
&D=
\left[
\begin{array}{cccc}
w-a & 0 & -c \gamma_{\bf k} & -e \gamma_{\bf k} \\
0 & w - b & -e \gamma_{\bf k} & -d \gamma_{\bf k} \\
c \gamma_{\bf k} & -e \gamma_{\bf k} & w + a & 0 \\
-e \gamma_{\bf k} & d \gamma_{\bf k} & 0 & w +b
\end{array}
\right]&
\nonumber \\ \nonumber \\
&N=\frac{1}{z}
\left[
\begin{array}{rrrr}
 c & -e &  c &  e \\
 e & -d &  e &  d \\
-c &  e & -c & -e \\
 e & -d &  e &  d
\end{array}
\right]&
\end{eqnarray}
with $w=\omega{\scriptstyle/}z$ and
\begin{eqnarray}\label{w16}
&&a=S_{\alpha}^{44}(S_{\alpha}^{22}-S_{\alpha}^{11})
\makebox[3.5mm][c]{,}
b=S_{\alpha}^{44}(S_{\alpha}^{11}-S_{\alpha}^{33})
\nonumber \\
&&c=S_{\alpha}^{21}S_{\alpha}^{34}
\makebox[3.5mm][c]{,}
d=-S_{\alpha}^{13}S_{\alpha}^{42}
\makebox[3.5mm][c]{,}
e=-S_{\alpha}^{13}S_{\alpha}^{34}=\sqrt{cd}
\;\;.
\end{eqnarray}
With this the longitudinal spin susceptibility of eqn. (\ref{w13}) is
obtained readily as
\begin{equation}\label{w17}
\chi^S_{\alpha\alpha}({\bf k},\omega) =
\frac{Z({\bf k},w){\scriptstyle/}z}{(w^2-w_1^2)(w^2-w_2^2)}
\;\;,
\end{equation}
where the weight factor $Z({\bf k},w)$ given by
\begin{eqnarray}\label{w18}
Z({\bf k},w)=&&2
(ac-bd -(c-d)^2\gamma_{\bf k})w^2
\nonumber \\
&&+(ad-bc)(ab+(ad-bc)\gamma_{\bf k})
\;\;,
\end{eqnarray}
and the excitation energies $\pm w_{1,2}({\bf k})$ are being set by
the roots of the biquadratic equation
\begin{eqnarray}\label{w19}
w^4+w^2((c-d)^2\gamma_{\bf k}-(a^2+b^2))
&&\nonumber \\
+a^2b^2-(ad-bc)^2\gamma_{\bf k}&&=0
\;\;.
\end{eqnarray}
In fig.(\ref{diszfacfig}) the dispersion as well as the weight
$R_{1,2}({\bf k})=
\chi^S_{\alpha\alpha}({\bf k},\omega)(\omega-\omega_{1,2}({\bf k})
|_{\omega=\omega_{1,2}({\bf k})}$
of the two po\-si\-tive-fre\-quen\-cy modes is depicted along a path in the
Brillouin zone (BZ) ranging from ${\bf k}=(1,1,1)$ to $(0,0,0)$ to
$(1,0,0)$ for various values of the anisotropy ratio $\xi/\eta$.
This figure clarifies the concluding issue aimed at in this brief
note, i.e. the observation of only a {\em single} excitation mode in
NdB$_6$. Based on the eigenvalues (\ref{g8}) {\em two} excitations of
comparable energy are expected in the Weiss field of the AFM state at
$\xi/\eta\ll 1$. However, fig. (\ref{diszfacfig}) shows that only a
single mode carries significant weight at small $\xi/\eta$. We have
shown this feature to remain valid for arbitrary range couplings
\cite{longvers}. Furthermore, in agreement with the spectrum of a
single-ion pseudo-spin ${\cal J}=3/2$, the system exhibits a
single-mode spin-wave like excitation at the isotropic point
$2\xi=\eta$. Only for intermediate anisotropy both modes show sizeable
weight at any given point in the BZ.

\subsection*{Conclusion}
In summary we have considered rare-earth compounds of cubic symmetry
with a $\Gamma_8$-quartet ground-state of the RE ions.
Particular emphasis has been put on the hexaboride NdB$_6$. Analyzing
the CEF splitting we have identified NdB$_6$ to be a genuine example of
a system with strongly coupled magnetic and quadrupolar degrees of freedom.

We have studied the CEF induced intrinsic magnetic anisotropy superimposed
onto an isotropic exchange interaction
revealing that NdB$_6$ should display magnetic anisotropy of a
different type, i.e. 'easy diagonal', as compared to Ce or
Yb compounds which show 'easy axis' anisotropy.

The magnetic anisotropy leads to a non-collinear ${\bf M}$ vs. ${\bf H}$
behavior and is tempting to speculate that angular-dependent magnetization
measurements on the corresponding RE cubic compounds, as well as diluted
systems, e.g, La$_{1-x}$Ce$_x$B$_6$, should be able to detect this behavior.

We have evaluated the magnetic excitations in the AFM state of an
'easy diagonal' type using a dyadic operator approach. For systems with
strong spin-quadrupolar coupling this method is superior \cite{longvers} to
less controlled pseudo-particle descriptions which are applicable to the
weak-coupling system CeB$_6$ \cite{ohkawa,ukf,sst} and are based on the
conventional ${\bbox \sigma}$-${\bbox \tau}$ Pauli-matrix representation
(\ref{g4}). In accordance with the number of
independent Pauli matrices (${\bbox \sigma}$ and ${\tau}$), we find two
branches of spin excitations. However, the spectral weights in the two
magnetic channels are very different in a strongly coupled spin-quadrupolar
system. In fact, in the $\xi=0$ limit one channel disappears completely.
This is reminiscent of the INS data on NdB$_6$ \cite{erkelens} which display
only one branch of spin excitations.
Although derived by a linearization of the EQMs we believe that our results
are quite robust against non-linear corrections since the spin-wave spectrum
in non-isotropic case is gapful. This should diminish the relevance of
quantum fluctuations.

Finally, regarding a direct comparison to experimental data we note that
NdB$_6$ displays a $[0,0,1/2]$ wave vector of the AFM modulation. This
requires the inclusion of longer-range exchange interactions which have
been neglected in this paper. These and the physical nature of the
branch unobservable by INS will be studied elsewhere \cite{longvers}.


\begin{thebibliography}{99}
\bibitem{refnote}
This has to be contrasted against claims of $[001]$ polarization in
\cite{erkelens} which deserve additional confirmation.
\end{thebibliography}
\end{document}